\documentstyle[12pt,epsf]{article}
\def\be{\begin{equation}}
\def\ee{\end{equation}}
\def\rg{renormalisation\ group}
\def\H{{\cal H}}
\def\fp{fixed\ point}
\def\ope{operator\ product\ expansion}
\def\ffrac#1#2{\textstyle{#1\over#2}\displaystyle}
\gdef\journal#1, #2, #3, 1#4#5#6{{\sl #1~}{\bf #2}, #3, 1#4#5#6}

\begin{document}
\title{Effect of Random Impurities on Fluctuation-Driven First Order
Transitions}
\author{John Cardy\\All Souls College and\\Department of Physics\\
Theoretical Physics\\1 Keble Road\\Oxford OX1 3NP, UK}
\maketitle

\begin{abstract}
We analyse the effect of quenched uncorrelated randomness coupling to the local
energy density of a model consisting of $N$ coupled two-dimensional Ising
models. For $N>2$ the pure model exhibits a fluctuation-driven 
first order transition, characterised by runaway renormalisation group 
behaviour.  We show that the addition of weak randomness acts to stabilise 
these flows, in such a way that the trajectories ultimately flow back towards
the \em pure \em decoupled Ising fixed point, with the usual critical
exponents $\alpha=0$, $\nu=1$, apart from logarithmic corrections.
We also show by examples that, in higher dimensions, such transitions
may either become continuous or remain first order in the presence of
randomness.
\end{abstract}
\newpage

The effect of quenched randomness coupling to the local energy density
of a system which, in its absence, undergoes a continuous phase
transition is well understood from the point of view of the
renormalisation group version of the Harris criterion\cite{Harris}.
When the specific
heat exponent $\alpha$ of the pure model is negative, weak randomness is
irrelevant from the \rg\ point of view, and the pure \fp\ is stable. On
the other hand, when $\alpha>0$ it is relevant, and, at least when the
cross-over exponent $\alpha$ is small, it may be argued that the
critical behaviour is controlled by a new, random, \fp\ close by. 

The effect on systems which undergo thermal first
order transitions is more dramatic. It was argued some time ago by
Imry and Wortis\cite{Imry} that, in two dimensions,
such systems should always exhibit a continuous
transition in the presence of such randomness. This is because the
random impurities couple to the local energy density in much the same
way that a random field couples to the local magnetisation in an Ising
system. In dimensions $d\leq2$, the Imry-Ma argument\cite{Imry-Ma} implies
that such random fields should destroy the ordered phases at low
temperature, and therefore also the first order phase boundary between
them. A similar argument, applied to randomness coupling to the local
energy density, then implies that a non-zero latent heat is impossible in two
dimensions in random systems whose pure versions exhibit such behaviour.
This argument has been rediscovered and put on a rigorous basis by
Aizenman and Wehr\cite{Aizenman}, and is supported by the phenomenological
and approximate renormalisation group arguments of Hui and
Berker\cite{Hui}. Monte Carlo work of Chen, Ferrenberg and
Landau\cite{Chen} on the $q=8$ state Potts model and of Domany 
and Wiseman\cite{Domany}
on the Ashkin-Teller and 4-state Potts models supports this
conclusion, and goes further: the continuous transition found by these
workers exhibits critical exponents which are consistent with those of
the pure Ising model, namely $\gamma/\nu\approx 1.75$,
$\beta/\nu\approx0.125$ and $\alpha\approx0$. An argument explaining
these findings has been put forward by Kardar, Stella, Sartoni and
Derrida\cite{Kardar}. They study the properties of an interface in the
$q$-state random-bond Potts model at low temperatures.
For $q\not=2$ this has a branching structure, but the authors argue, on
the basis of simplified recursion relations which are exact on a
hierarchical lattice, that the critical behaviour where the 
interfacial free energy vanishes is governed by a zero-temperature 
\fp\ (as in the random field problem), and the Widom exponent $\mu$
which governs the vanishing of the surface tension is independent of $q$
for sufficiently large $q$, being numerically consistent with the Ising
value $\mu=1$. 

The first order transition in the pure $q$-state Potts model is of mean
field type, that is it is already predicted on the basis of mean field
theory. Such `strong' first order transitions are
described within the \rg\ by zero-temperature, discontinuity
\fp s, characterised by a relevant 
\rg\ eigenvalue $y=d$ whose scaling field couples to the local energy
density. Quenched randomness coupling to this has
eigenvalue $d-2(d-y)=d$, and is therefore also strongly
relevant. It is thus difficult to treat the effects of such
randomness systematically within a controlled \rg\ calculation.

In this paper, by contrast, we study the effects of quenched randomness
coupling to the local energy density on systems whose pure versions
exhibit \em fluctuation-driven \em first order transitions. These are
transitions which are expected to be continuous on the basis of a mean
field analysis, but which are driven first order by the fluctuation
effects. In terms of the \rg, they are often characterised by so-called
runaway behaviour, that is, the \rg\ trajectories move out of the region
in which the original perturbative calculation is valid. That, in
itself, does not guarantee that the system in question undergoes a first
order transition, but often it is possible to argue that the
trajectories then move into a region where mean field theory is applicable,
and which may then predict a first order transition.
The Imry-Wortis argument\cite{Imry,Aizenman} should, of course,
apply equally well to systems exhibiting
fluctuation-driven first order transitions.
However, the advantage of studying these
from the \rg\ point of view is that it is
possible to analyse them within a controlled perturbative
scheme and to elucidate the nature of the \fp\ which governs the
continuous critical behaviour of the random system.

\begin{sloppypar}
A simple example of a two-dimensional system which exhibits a
fluctuation-driven first order transition is that of $N$ Ising models coupled
through their local energy densities. Microscopically this may be
represented in terms of a lattice model with $N$ Ising spins
$(s_1(r),\ldots,s_N(r))$ at each site $r$ of the lattice. The reduced
hamiltonian is
\be
\label{NI}
H=-K\sum_i\sum_{r,r'}s_i(r)s_i(r')-g\sum_{i\not=j}\sum_{r,r'}s_i(r)s_i(r')
s_j(r)s_j(r'),
\ee
where the sums over $(r,r')$ are over nearest neighbour sites. Such a
model is self-dual on the square lattice, 
so that the critical coupling $K_c$ may be found exactly. 
In the absence of randomness,
the \rg\ equations on the critical surface have the form 
\be
\label{dgdl}
dg/d\ell=(N-2)g^2+O(g^3).
\ee
For $N=2$, this vanishes, as expected since this case corresponds to the
Ashkin-Teller model which exhibits a line of fixed points labelled by
$g$.\cite{ATmodel} When $N>2$, however, initially small positive values of $g$
flow out of the region of validity of the perturbative equation
(\ref{dgdl}). When (\ref{NI}) is analysed within mean field theory, 
the quartic term in the free energy remains positive if $g$ is
small, indicating a continuous transition, but, for sufficiently large
$g$, it changes
sign so that the mean field transition becomes first order. 
Since the \rg\ indicates that, no matter what the initial value of $g$,  
it should ultimately renormalise into this region, it implies that the
transition should be first order for \em all \em $g$, and is therefore
of a fluctuation-driven nature for small $g$. 
In fact, on the critical surface, this model when expressed in terms of
Ising fermions is nothing but the Gross-Neveu model\cite{Gross}, 
which is believed
to be massive for $N>2$, corresponding to a finite correlation length.
\end{sloppypar}

We now consider adding quenched randomness which couples to the local
energy density. This may be done in a variety of ways, but, in order to
focus on the universal properties of such a coupling, let us first
rewrite (\ref{NI}) in a continuum notation in terms of the local energy
density $E_i(r)$ of each Ising model. The hamiltonian density, close to
the critical point of the pure system, may then be written
\be
\H=\H_c+t\sum_iE_i-g\sum_{i\not=j}E_iE_j,
\ee
where $H_c$ is the \fp\ hamiltonian and
$t$ is the temperature deviation from the critical point. Quenched randomness
is now added by allowing $t\to t+\delta t(r)$, where $\overline{\delta
t(r)}=0$ and $\overline{\delta t(r)\delta t(r')}=\Delta\delta(r-r')$.
Introducing $n$ replicas $a=1,\ldots,n$
and averaging over a Gaussian distribution, the
replicated hamiltonian density is
\be\label{Hrep}
\H=\sum_a\H_c^a+t\sum_{i,a}E_i^a-g\sum_{i\not=j,a}E_i^aE_j^a
-\Delta\sum_{i,j,a,b}E_i^aE_j^b.
\ee
Note that each term has a well-defined behaviour under the duality
operation under which $E_i^a$ reverses sign. The self-dual critical
point therefore remains at $t=0$ in this parametrisation. It is possible
to consider replica-coupling terms which break this duality symmetry,
but they are all irrelevant close to the pure decoupled \fp.

The perturbative \rg\ equations for the couplings follow using standard
methods. In general for a perturbed hamiltonian density of
the form $\H=\H_c+\sum_ig_i\Phi_i$, they have the form\cite{Book}
\be
dg_k/d\ell=y_kg_k-\sum_{i,j}c_{ijk}g_ig_j+O(g^3),
\ee
where $y_i$ is the eigenvalue at the unperturbed \fp, and $c_{ijk}$
is the coefficient of $\Phi_k$ in the \ope\ of $\Phi_i$ with $\Phi_j$. 
In the present case, these are very easy to work out. Both the
interaction terms in (\ref{Hrep}) have a similar form, and are in fact special
cases of a very general model of $Nn$ interacting Ising models, with a
hamiltonian density
\be
\H=\H_c+\sum_{p\not=q=1}^{Nn}G_{pq}E_pE_q.
\ee
The terms with $p=q$ are excluded since the \ope\ of $E_p$ with itself
in the Ising model yields only the trivial identity operator. 
Normalising the energy density
so that $E_p\cdot E_{p'}=\delta_{pp'}$, the required terms in the 
\ope\ are then
\be
(E_pE_q)\cdot(E_{p'}E_{q'})=\delta_{pp'}(E_qE_{q'})+{\rm
permutations}+\cdots,
\ee
from which follow the general
\rg\ equations
\be
dG_{pq}/d\ell=-4(1-\delta_{pq})\sum_rG_{pr}G_{rq}+O(G^3).
\ee

\begin{figure}[t]
\centerline{
\epsfxsize=4.5in
\epsfbox{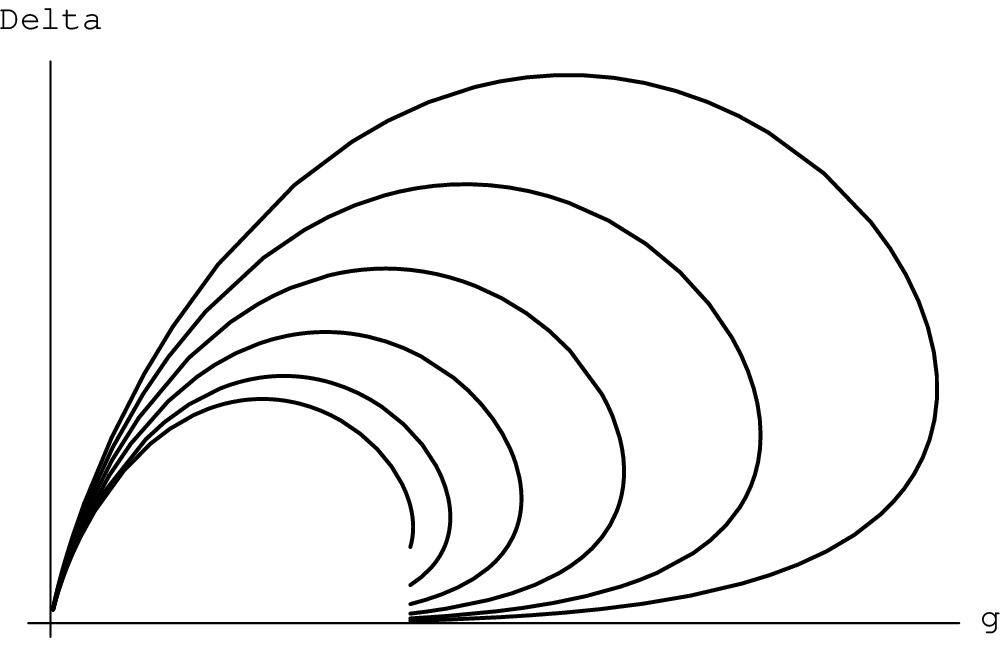}}
\caption{RG flows on the critical surface for the case $N=3$. Several
trajectories are shown, for different initial values of the randomness $\Delta$
at a fixed value of the coupling $g$ between the Ising models. Although
they initially flow towards strong coupling, they eventually curl back
towards the pure Ising fixed point.}
\label{fig1}
\end{figure}

Specialising these to the case at hand, we then find, in the limit
$n\to0$, the flow equations
\begin{eqnarray}
dg/d\ell&=&4(N-2)g^2-8g\Delta+\cdots\label{flows1}\\
d\Delta/d\ell&=& -8\Delta^2+8(N-1)g\Delta+\cdots\label{flows2}\\
dt/d\ell&=& t\big(1-4\Delta+4(N-1)g\big)+\cdots.\label{flows3}
\end{eqnarray}
The last equation in fact follows from the second by the \rg\ version
of the Harris criterion\cite{Book2}.
For the case $N=2$ these equations are equivalent to those found for the random
Ashkin-Teller model by Dotsenko and Dotsenko\cite{D2}.
Remarkably, the flows in the critical surface obtained by solving these 
equations may be found in closed form for general $N$:
\be
g={\rm const.}(\Delta/g)^{(N-2)/N}\,e^{-2\Delta/Ng}.
\ee
These are shown in Fig.~\ref{fig1} for the case $N=3$. 
For $g=0$ we find that the randomness is marginally
irrelevant, consistent with the well-known case of the random Ising
model\cite{DD,Ludwig1}. In the absence of randomness, the flows for
$g>0$ run away to the first order region, as discussed above.
However, for any non-zero
randomness the trajectories eventually
curl around and approach the fixed point corresponding to $N$ decoupled
pure Ising models. Of course equations (\ref{flows1}-\ref{flows3}) 
are strictly valid
only inside the perturbative region where the initial values of the
parameters are small, but it is reasonable to expect that the topology of
the \rg\ flows should persist at least in some finite region around the
origin. This topology has two important consequences: first, as
dictated by the Imry-Wortis argument, the transition has
become continuous,
and secondly, the asymptotic critical behaviour is that of the pure
Ising model, similar to those cases discussed earlier. 
In fact, by integrating equation (\ref{flows3}) for $t$ it may be
shown that the specific heat has a singularity of the form
$A\ln\ln(1/t)$, just as for the random bond Ising model\cite{DD}, but with
an amplitude $A\propto\Delta^{-(N-2)/N}$.
Flows of the type shown in Fig.~\ref{fig1} are very unusual as they
violate the $c$-theorem\cite{Zam}. Of course, that this can happen is a
consequence of the $n\to0$ replica limit. 

It is instructive to extend the above analysis to dimensions $d=2+\epsilon$,
since the Imry-Wortis argument leads to no definite conclusion in that
case. The perturbative \rg\ equations become
\begin{eqnarray}
dg/d\ell&=&\alpha_pg+\big(4(N-2)+2b^2\big)g^2-(8-4b^2)g\Delta+\cdots\\
d\Delta/d\ell&=&\alpha_p\Delta-(8-2b^2)\Delta^2+8(N-1)g\Delta+\cdots\\
dt/d\ell&=&t\big(d/(2-\alpha_p)-(4-2b^2)\Delta+4(N-1)g\big)+\cdots.
\end{eqnarray}
The linear terms are a consequence of the fact that the specific heat
exponent $\alpha_p$ of the pure model no longer vanishes: this
determines the eigenvalues of $g$ and $\Delta$ at this \fp\ according to
the Harris criterion. The parameter $b$ is the \ope\ coefficient
appearing in $E_p\cdot E_q=\delta_{pq}+b\delta_{pq}E_p+\cdots$. With the
energy density normalised in this way, it is universal, depending only
on $d$, but it vanishes for $d=2$ as a consequence of duality, and is
therefore presumably small just above two dimensions. (In $d=4$, at the
Gaussian \fp, $b=2\surd2$.) When $g=0$, the randomness is now relevant,
and the critical behaviour is controlled by a non-trivial random
\fp\ at $\Delta=O(\alpha_p)$.\cite{Book2} However, when $b\not=0$, $g$
is in fact relevant at this \fp, and there exists another, more stable
\fp, which, for $b\ll1$, is located at $g\approx b^2\alpha_p/16N$,
$\Delta\approx\alpha_p/8$. 
When $g>0$ initially, the
trajectories move towards larger values of $g$ before 
eventually curling around to finish at this new coupled random
\fp. Therefore this gives an example of a fluctuation-driven first order
transition in $d>2$ dimensions, which is converted to a continuous
transistion, as in $d=2$. However, now the critical behaviour is
controlled by a new, random, \fp. Such a \fp, if it persists as far as
$\epsilon=1$, would describe the random Ashkin-Teller model in three
dimensions, at least for small values of $g$.

The above calculation breaks down near four dimensions (if not before),
due to the proximity of the Gaussian \fp. As a further example of what can
happen to a fluctuation-driven first order transition in $4-\epsilon$
dimensions, consider the well-known problem of the $O(N)$, or
$N$-vector, model, with cubic symmetry breaking\cite{cubic4d}.
This model has $N$-component continuous spins $S_i(r)$, and the
replicated hamiltonian density is
\be
\H=t\sum_{i,a}\big(S_i^a\big)^2+u\sum_{i,j,a}\big(S_i^a\big)^2
\big(S_j^a\big)^2+v\sum_{i,a}\big(S_i^a\big)^4
-\Delta\sum_{i,j,a\not=b}\big(S_i^a\big)^2\big(S_j^b\big)^2.
\ee
For $u\ll v$ this may be viewed as a continuous spin version of
(\ref{Hrep}), with $g=-u$. However, in the absence of randomness
this model also possesses an $O(n)$ \fp\ (which is absent for $d=2$) and
a cubic \fp\ where both $u$ and $v$ are non-zero. 
The perturbative \rg\ equations may be found from the \ope\ as above:
\begin{eqnarray}
du/d\ell&=&\epsilon u-8(N+8)u^2+8N\Delta^2-48uv+\cdots\\
dv/d\ell&=&\epsilon v-72v^2-96uv+\cdots\\
d\Delta/d\ell&=&\epsilon\Delta-16(N-2)\Delta^2-16(N+2)u\Delta
-48v\Delta+\cdots.
\end{eqnarray}
When $\Delta=0$ these exhibit runaway behaviour to the first order
region where $u$ and $v$ are large and negative, if the initial value
of $v$ is sufficiently negative (for $N>4$, when the cubic \fp\ is in
the quadrant with $v>0$, this requires only that $u>0$ and $v<0$ initially).
However, since $\Delta$ does not enter the flow equation for $v$, 
this catastrophe still occurs in the presence of randomness. We conclude
that quenched randomness does not change the order of the transition in
this case. This is, of course, quite consistent with the Imry-Wortis
argument, which does not rule out either behaviour for $d>2$. 

The case where $u<0$ and $v>0$ corresponds to the same example of
$N$ coupled Ising models as before, this time near four dimensions.
Solving the \rg\ equations for $v$ near the decoupled Ising \fp, we find
(using $g$ rather than $u$)
\begin{eqnarray}
dg/d\ell&=& \ffrac13\epsilon g+8Ng^2+32g\Delta+\cdots\label{aa}\\
d\Delta/d\ell&=&\ffrac13\epsilon \Delta+16(N-2)g\Delta+\cdots.\label{bb}
\end{eqnarray}
When $g=0$, these equations have no perturbative \fp, despite the fact
that $\Delta$ is relevant. This is the well-known problem of the random
Ising model near $d=4$, and it is cured in a higher
order calculation\cite{randomising4}, when a term $O(\Delta^3)$ appears on the 
right hand side of (\ref{bb}), giving an $O(\epsilon^{1/2})$ \fp.
However, it may be seen from the structure of the other terms in
(\ref{aa},\ref{bb}) that this cannot cure 
the runaway behaviour which occurs once
the coupling $g$ is initially non-zero. We conclude that the
fluctuation-driven first order transition probably persists in this
case.

As a final example, we may quote the case of the complex $O(N)$ model
near four dimensions, coupled to a long-range $U(1)$ gauge field,  known
as the Abelian Higgs model for the case $N=1$. This
was argued long ago to undergo a fluctuation-driven first order
transition near four dimensions for sufficiently small $N$.\cite{HLM}
The effect of quenched random impurities was studied by Boyanovsky
and Cardy\cite{Boyanovsky}, who found that for sufficiently weak
randomness the first order nature of the transition persists, while for
stronger randomness
the trajectories spiral in towards a new random \fp,
corresponding to a continuous transition.

To summarise, we have given examples of how quenched randomness
coupling to the local energy density converts a fluctuation-driven first
order transition into a continuous one for $d=2$, consistent with the
Imry-Wortis argument, and of how this may or may not happen when $d>2$.

We conclude with a discussion of the conjecture that all
random critical behaviour in two dimensions is Ising-like.
This is based on numerical results on the random 
Ising model\cite{Dots,Wang1,Wang2},
the Ashkin-Teller model and the 4-state Potts
model\cite{Domany}\footnote{
There is also claimed experimental evidence for the 4-state
Potts model\cite{expt} -- however the randomness discussed there would
appear to favour one sublattice rather than another, and
therefore should couple to the order parameter, corresponding to the 
random \em field \em problem.}
(all of which exhibit continuous transitions in the
absence of randomness), and the 8-state Potts model (which is first
order in its pure version.) It is backed up by the interface arguments
of Kardar et al.\cite{Kardar}, which suggest that the Widom exponent for
the random $q$-state Potts model is independent of $q$ for sufficiently
large $q$. (A similar lack of dependence on $q$ has been argued for in
the case of random Potts spin chains\cite{Majumdar}; 
however, their critical behaviour
is rather different in nature from that of the present case.)

The conjecture in the case of the Ising model and the Ashkin-Teller model
(close to the decoupling point) agrees with the results of a
perturbative \rg\ analysis\cite{D2}, and with our analysis above: the
\rg\ trajectories curl around and end up at the Ising \fp, giving Ising
exponents, but with logarithmic (or log-log) modifications.
However, a similar analysis\cite{Ludwig,Dots2}
applied to the random $q$-state Potts model
for $q>2$ indicates the existence of a new random \fp\ whose critical
exponents are not Ising-like, but depend on $q$. This analysis is valid
only when $q-2$ is small, but is consistent with earlier \rg\
results\cite{DG} for $q=3$. 

We have not been able to resolve this discrepancy, but would venture a
few remarks which, in fact, may seem to confuse the situation further:
\begin{enumerate}
\item  The perturbative \rg\ arguments work with a distribution of
randomness which is \em self-dual \em, corresponding on the lattice, for
example, to an equal distribution of strong and weak bonds of strengths
$K$ and $K^*$ which are dual to each other. Within the perturbative
scheme, this is justified, since it may be argued that weak randomness
which violates self-duality is irrelevant in the \rg\ sense. 
\item However,
the interfacial analysis of Kardar et al.\cite{Kardar}, which treats
horizontal and vertical bonds on a quite different footing, cannot, by its
nature, respect the duality properties of the model. Indeed, these
authors find it necessary to include negative bonds in the model to
access their zero-temperature \fp, which are excluded in any self-dual
formulation of the problem. The only zero-temperature \fp\ in the
self-dual random model is the percolation point. 
\item This leads to the picture that the critical behaviour controlled
by a zero-temperature \fp, discussed by Kardar et al.\cite{Kardar}, and
that found in the perturbative \rg\ of Ludwig\cite{Ludwig}, are simply
different and correspond to strong
non-self-dual randomness and to weak self-dual randomness respectively.
However, the numerical results for the $q=4$ and $q=8$ Potts models,
which appear to find Ising-like exponents independent of $q$, use
\em self-dual \em randomness in order to locate the critical point precisely.
They also consider different strengths of randomness, with no
appreciable difference in their results. 
\item One would expect critical behaviour controlled by a
zero-temperature \fp\ to exhibit hyperscaling violation, as in the
random field problem. However, the exponents found in the numerical work
for $q=8$ are consistent with hyperscaling, that is, with a
conventional, finite-temperature fixed point as found in the
perturbative \rg\ approach.
\end{enumerate}

Whatever the resolution of this problem, it cannot be that all the
universal properties of the random $q$-state Potts model are independent
of $q$, even if the exponents are. This is because this critical point
separates a $q$-fold degenerate ordered phase from a nondegenerate
disordered phase, and this degeneracy must reflect itself in the
fluctuation contribution to the free energy near the critical point,
even if the exponents are Ising-like. This may be seen in the example of
$N$ coupled Ising models discussed in this paper: although the relevant
critical \fp\ is Ising-like, it in fact corresponds to $N$ decoupled
Ising models, not just one. This will reflect itself  
in universal amplitude ratios which
involve the free energy. However, because of the expected logarithmic
corrections, these may be difficult to analyse from numerical data.
A cleaner test should be through the value of the effective central
charge (which measures the finite-size scaling behaviour of the quenched
free energy\cite{BCN,Aff}). In our example, this is $c=\frac12N$, 
and does depend on $N$. It would be very interesting to compute this for
the random $q$-state Potts model. To how many decoupled Ising models
does the random $q$-state Potts model correspond at criticality, if
indeed its behaviour is Ising-like?

The author acknowledges useful discussions with E.~Domany, 
M.~Lassig, S.~Majumdar and S.~Wiseman, and thanks the Weizman Institute,
where this work was initiated, for its hospitality. 
This work was supported in part by EPSRC grant GR/J78327.


\begin{thebibliography}{99}
%
\bibitem{Harris}Harris A B, \journal J. Phys. C, 7, 1671, 1974.
%
\bibitem{Imry}Imry Y and Wortis M, \journal Phys. Rev. B, 19, 3581,
1979.
%
\bibitem{Imry-Ma}Imry Y and Ma S-K, \journal Phys. Rev. Lett., 35, 1399,
1975.
%
\bibitem{Aizenman}Aizenman A and Wehr J, \journal Phys. Rev. Lett., 62,
2503, 1989.
%
\bibitem{Hui}Hui K and Berker N, \journal Phys. Rev. Lett., 62, 2507, 1989;
{\bf 63}, 2433(E), 1989.
%
\bibitem{Chen}Chen S, Ferrenberg A M and Landau D P, \journal Phys. Rev.
E, 52, 1377, 1995.
%
\bibitem{Domany}Domany E and Wiseman S, \journal Phys. Rev. E, 51, 
3074, 1995.
%
\bibitem{Kardar}Kardar M, Stella A L, Sartoni G and Derrida B, \journal
Phys. Rev. E, 52, R1269, 1995.
%
\bibitem{ATmodel}Kadanoff L P, \journal Phys. Rev. B, 22, 1405, 1980.
%
\bibitem{Gross} Gross D and Neveu A, \journal Phys. Rev. D, 10, 3235,
1974.
%
\bibitem{Book}Cardy J L, {\sl Scaling and Renormalization in Statistical
Physics}, Ch.5 (Cambridge University Press, 1996).
%
\bibitem{Book2}\em ibid.\em, Ch.8.
%
\bibitem{Majumdar}Senthil T and Majumdar S, preprint.
%
\bibitem{D2}Dotsenko V S and Dotsenko Vl S, \journal J. Phys. A, 18, L241,
1985.
%
\bibitem{DD}Dotsenko V S and Dotsenko Vl S, \journal Adv. Phys., 32, 129,
1983. 
%
\bibitem{Ludwig1} Ludwig A W W, \journal Nucl. Phys. B, 330, 639, 1990.
%
\bibitem{Zam}Zamolodchikov A B, \journal Pis'ma Zh. Eksp. Teor. Fiz.,
43, 565, 1986. [\journal JETP Lett., 43, 730, 1986.]
%
\bibitem{cubic4d}Aharony A, in {\sl Phase Transitions and Critical
Phenomena}, vol.6, p.394, C.~Domb and M.~S.~Green, eds., (Academic,
1976.)
%
\bibitem{randomising4}Lubensky T, \journal Phys. Rev. B, 11, 3573, 1975.
%
\bibitem{HLM}Halperin B, Lubensky T and Ma S-K, \journal Phys. Rev.
Lett., 32, 292, 1974.
%
\bibitem{Boyanovsky}Boyanovsky D and Cardy J L,  \journal Phys. Rev. B,
25, 7058, 1982.
%
\bibitem{Dots}Andreichenko V B, Dotsenko Vl S, Selke W and
Wang J-S, \journal Nucl. Phys. B, 344, 531 1990.  
%
\bibitem{Wang1}Wang J-S, Selke W,
Dotsenko Vl S and Andreichenko V B,
\journal Europhys. Lett., 11, 301, 1990.
%
\bibitem{Wang2}Wang J-S, Selke W, Dotsenko Vl s and Andreichenko V B,
\journal Physica A, 164, 221 1990.
%
\bibitem{expt}Schwenger L, Budde K, Voges C and Pfn\"ur H, 
\journal Phys. Rev. Lett., 73, 296, 1994.
%
\bibitem{DG}Derrida B and Gardner E, \journal J. Phys. A, 17, 3223,
1984.
%
\bibitem{Ludwig}Ludwig A W W, \journal Nucl. Phys. B, 285, 97, 1985.
%
\bibitem{Dots2}Dotsenko Vl, Picco M and Pujol P, {\sl Nucl. Phys.}, to
appear.
%
\bibitem{BCN}Bl\"ote H, Cardy J L and Nightingale M P, \journal
Phys. Rev. Lett., 56, 742, 1986.
%
\bibitem{Aff}Affleck I, \journal Phys. Rev. Lett., 56, 746, 1986.










\end{thebibliography}
\end{document}